\documentstyle[aps,preprint,tighten,epsf]{revtex}

\begin{document}

\preprint{\vbox{\hspace{-0.25cm}Bonn TK-95-33 \hfill Submitted Physical Review
    C}}
\title{$K^0$ form factor and charge radius in a covariant Salpeter model}

\author{Wolfgang\,I.\,Giersche\footnote{E-mail: wolfgang@itkp.uni-bonn.de},
  Claus\,R.\,M\"unz\footnote{E-mail: muenz@itkp.uni-bonn.de}}
\address{Institut f\"ur Theoretische Kernphysik,
  Universit\"at Bonn, Nu{\ss}allee 14-16, \\ D-53115 Bonn, Germany}
\date{\today}

\maketitle

  \begin{abstract}
    The electromagnetic form factor for the $K^0$ is calculated in a
    covariant formulation of the Salpeter equation for $q\bar q$ - bound
states,
    which has been presented recently for the mass spectrum, decay properties
    and form factors of the light pseudoscalar and vector mesons. The $K^0$
    charge radius dependence on the difference between strange and down
    constituent quark mass is discussed.
  \end{abstract}

\section{Introduction} \label{I}
It has been proposed by Magahiz et al. \cite{Ma95} to observe the reaction
$p(e,e'K^0)\Sigma^+$ with the CLAS Large Acceptance Spectrometer at CEBAF to
gain insight into strangeness electroproduction of nuclei. If the longitudinal
and transverse part of the differential cross section could be separated, the
t-channel reaction would allow the measurement of the $K^0$ electromagnetic
form factor which due to the mass difference between the strange and down
quarks does non vanish.  In view of this proposed experiment predictions for
the $K^0$ form factor have been recently published and it commonly turned out
that due to the accessible values of momentum transfer up to few
${\mbox{(GeV)}}^2$ a covariant description of the underlying dynamics is
mandatory.  Cardarelli et al. investigated a relativistic constituent quark
model based on the light front formalism \cite{Ca95}. Therein they made use of
an interaction kernel motivated by an effective $q\bar q$ - Hamiltonian which
has been developed by Godfrey and Isgur \cite{IG85}.  In another paper
Buck, Williams and Ito \cite{BWI95} calculated the $\pi$ and $K$ form factors
by employing a model described in \cite{IBG92}, with a separable ansatz
including symmetry breaking effects.

In two previous papers \cite{MR94a,MR94b} we presented a covariant quark model
based on the Salpeter equation and used it to compute (transition) form factors
between the light pseudoscalar and vector mesons \cite{MR95}. This brief report
shall serve as an extension of the latter to the neutral strange meson.

\section{The Model}
Starting from the Bethe-Salpeter equation, we use a $q\bar q$ - interaction
assumed instantaneous in the rest frame of the bound state and free effective
quark propagators to arrive at the Salpeter equation, which is expressed as an
eigenvalue problem for the bound state mass and solved numerically
\cite{MR94a}. In addition to the calculation of mass spectra we have presented
there a method to reconstruct the four-dimensional Bethe-Salpeter amplitude
from the equal-time Salpeter amplitudes.

In our model the interaction consists of a confining potential which is
linearly rising in coordinate space, and an instanton-induced interaction
derived by \mbox{'t Hooft} (see \cite{MR94b} and references therein) as a
possible solution of the U$_A$(1)-problem.  We would like to emphasize that the
potential and mass parameters used in our model have been fixed in \cite{MR94b}
to obtain a reasonable agreement with the experimental mass spectrum of the low
lying pseudoscalar and vector mesons as well as the leptonic $\pi$ and $\rho$
decay widths. The electromagnetic form factors as well as the other decay
widths (e.g. $M \rightarrow M'\, \gamma$) are consistently obtained by
employing the Mandelstam formalism to the formerly calculated Bethe-Salpeter
amplitudes. No additional parameter is used to calculate the current matrix
elements. In lowest order we find for the electromagnetic current coupling to
the quark:
\begin{eqnarray}
  \label{current}
  \langle P' | \, j^{(1)}_{\mu}(x) \, | \, P \, \rangle =
  -e_1 \int \!\! \frac{d^4p}{(2\pi)^4}\, \mbox{tr} \!
  \left\{
  \left[(i \partial \!\!\! / \, -m_2)\bar \chi_{P'}(x,y) \right] \gamma_{\mu}
  \chi_P(x,y)
  \right\}
\end{eqnarray}
which is formally analogous to the results obtained by Buck et al.
\cite{BWI95} except for the inner structure of the amputated Bethe-Salpeter
amplitude, which in our model contains in general eight amplitudes for fixed
spin and parity \cite{MR95} and thus reflects the full Dirac structure of the
$q\bar q$ - system.

\section{Results: The $K^0$ electromagnetic form factor and charge radius}

\begin{figure}[htb]
  \begin{center}
    \leavevmode
    \epsfxsize=0.7\textwidth
    \epsffile{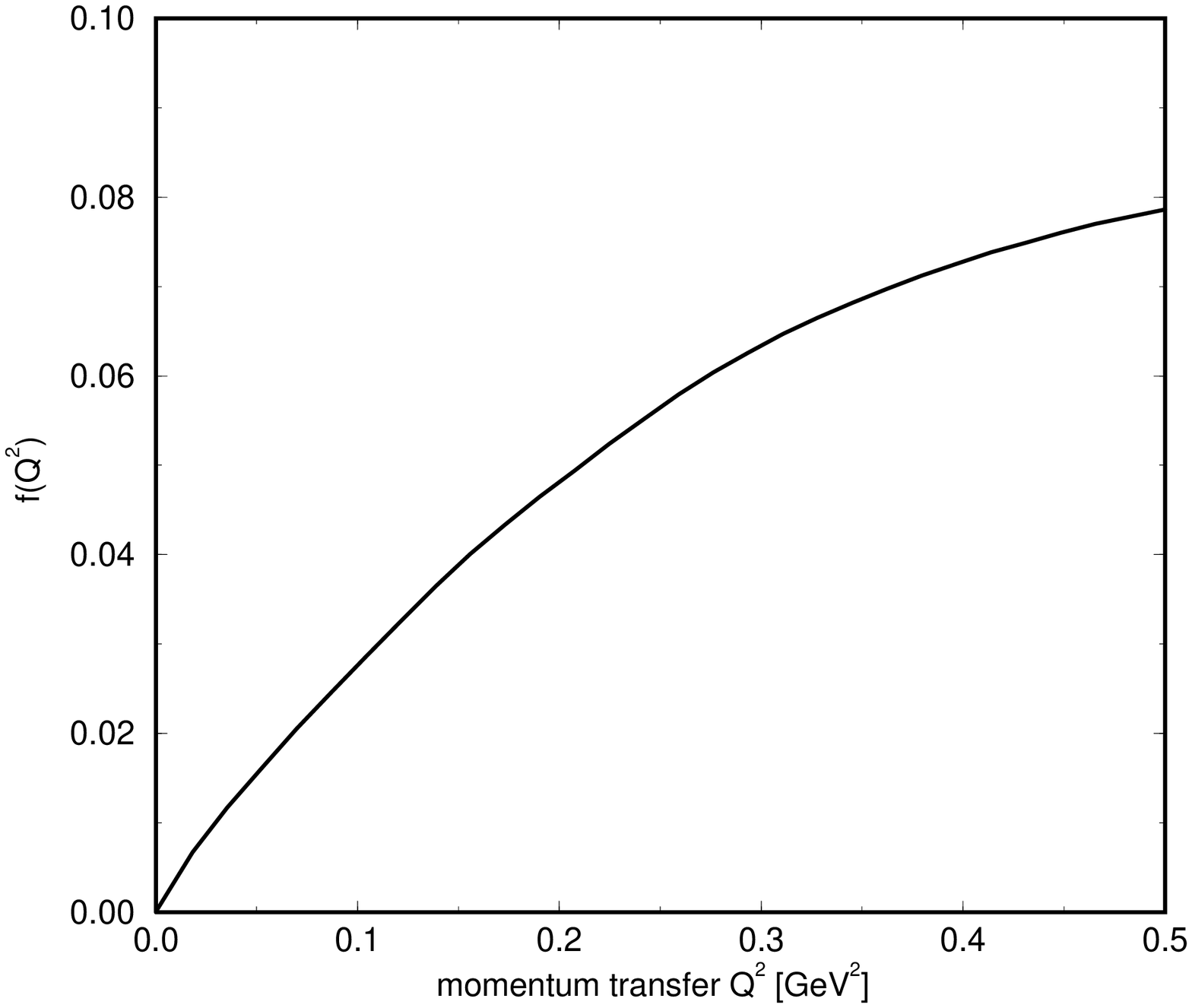}\vspace{-0.3cm}
    \caption{
      The $K^0$ form factor at small momentum transfer}
    \label{Abb1}
  \end{center}
\end{figure}

\begin{figure}[htb]
  \begin{center}
    \leavevmode
    \epsfxsize=0.7\textwidth
    \epsffile{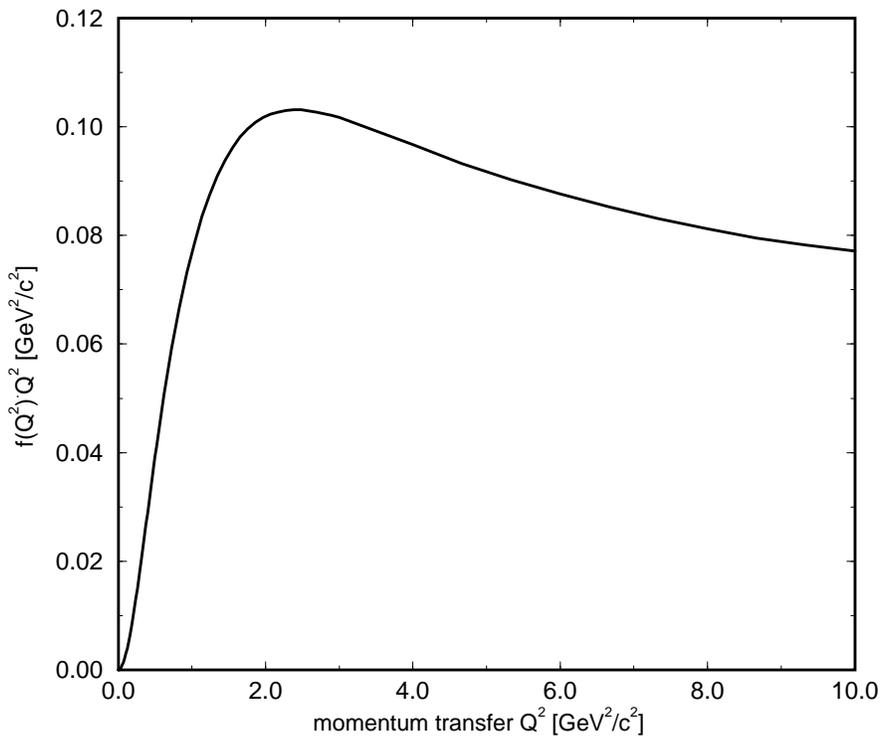}\vspace{-0.3cm}
    \caption{The $K^0$ form factor times $Q^2$ at large momentum transfer}
    \label{Abb2}
  \end{center}
\end{figure}

Our results for the $K^0$ form factor $f(Q^2)$ are shown in Fig. \ref{Abb1} and
\ref{Abb2} for small and large momentum transfer (in Fig. \ref{Abb2} we plotted
$Q^2\cdot f(Q^2)$). Our calculation agrees remarkably well with the prediction
of
Buck et al.~\cite{BWI95}, where the parameter have been fixed to reproduce the
$\pi^+$ and $K^+$ charge radii and decay constants. However, our maximum of
$Q^2\cdot f(Q^2)$ appears at a smaller momentum transfer of approximately 2
GeV$^2$.

As our calculation has been performed in the framework of a covariant quark
model which includes confinement, and therefore is able to describe not only
the masses and decay properties of the pseudoscalars but also of the vector
mesons, the $K^0$ form factor calculation is put on a more general basis than
in the work of Buck et al.~\cite{BWI95}, although our results do not differ
significantly.

\begin{figure}[htb]
  \begin{center}
    \leavevmode
    \epsfxsize=0.7\textwidth
    \epsffile{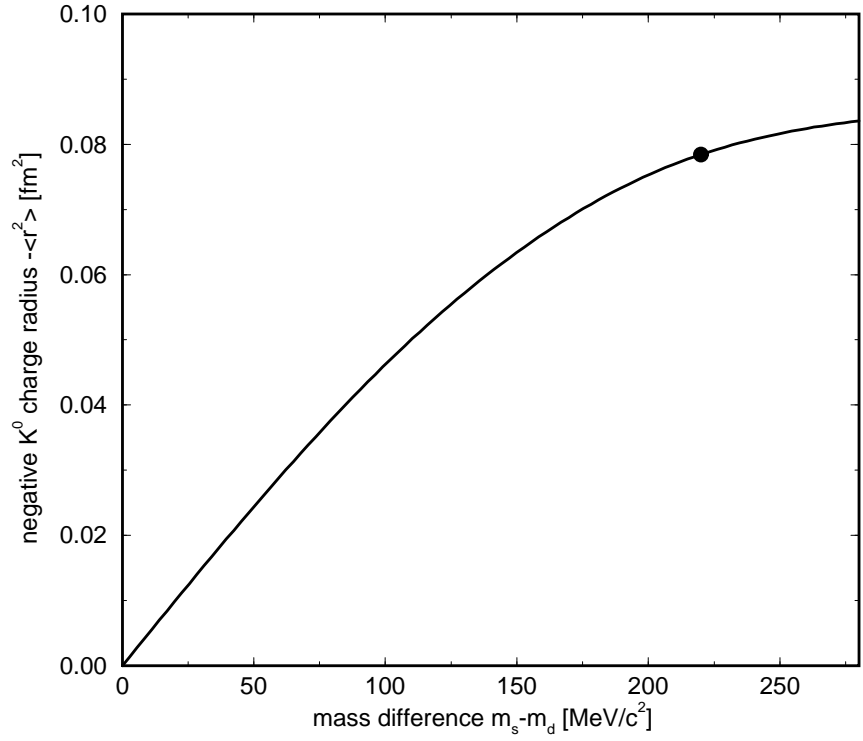}
    \caption{ The mean squared charge radius as a function of the
      difference between strange and down constituent quark mass. The dot
      indicates the prediction of our original model from
      \protect\cite{MR94b}.}
    \label{Abb3}
  \end{center}
\end{figure}

The $K^0$ charge radius, as has been discussed e.g.\ in \cite{Bur95}, is most
sensitive to the mass difference between the strange and down quark mass. We
have estimated the charge radius by a least-square fit of a quadratic function
to our form factor below 0.1 GeV$^2$ and studied its dependence on the
differences of the constituent quark masses $m_s - m_d$ keeping the sum of them
fixed to our original value $m_d + m_s = 170\mbox{MeV} + 390 \mbox{MeV} =
560\mbox{MeV}$~\cite{MR94b}. The results plotted in Fig. \ref{Abb3} indeed
shows a strong dependence on the quark mass difference, as long as it is
smaller than 250 MeV.

An experimental measurement of the $K^0$ charge radius therefore would be an
interesting opportunity to determine the difference between strange and
nonstrange constituent quark mass, alternatively to estimates from purely
spectroscopic quark model calculations.

\end{document}